\input harvmac.tex

\noblackbox
\lref\phases{E. Witten, ``Phases of N=2 Theories in Two Dimensions'',
{\it Nucl. Phys.} {\bf B}403 (1993) 159, hep-th/9301042}
\lref\dk{J. Distler and S. Kachru, ``(0,2) Landau-Ginzburg Theory'',
{\it Nucl. Phys.} {\bf B}413 (1994) 213, hep-th/9309110.}
\lref\edcomments{E. Witten, ``Some Comments on String
Dynamics'', hep-th/9507121.}
\lref\berkooz{M. Berkooz and R. Leigh, 
``A D=4 N=1 Orbifold of Type I Strings'', hep-th/9605049 and to appear.}
\lref\IT{K. Intriligator and S. Thomas, ``Supersymmetry
Breaking on Quantum Moduli Spaces'', hep-th/9603158.}
\lref\ganhan{O. Ganor and A. Hanany, ``Small E(8) Instantons
and Tensionless Noncritical Strings'', hep-th/9602120.} 
\lref\seiwit{N Seiberg and E Witten, ``Comments on String
Dynamics in Six Dimensions'', {\it Nucl. Phys.} {\bf B}471 (1996) 121, 
hep-th/9603003.}
\lref\bsv{M. Bershadsky, V. Sadov, and C. Vafa, ``D strings on D manifolds'',
{\it Nucl. Phys.} {\bf B}463 (1996) 398, hep-th/9510225.}
\lref\dsww{M. Dine, N. Seiberg, X. G. Wen, E. Witten,
``Nonperturbative Effects on the String Worldsheet'', 
{\it Nucl. Phys.} {\bf B}278 (1986) 769; {\it Nucl. Phys.} {\bf B}289
(1987) 319. }
\lref\oldwit{E. Witten, ``New Issues for Manifolds of $SU(3)$ Holonomy,''
Nucl. Phys. {\bf B}268 (1986) 79.}
\lref\witsup{E. Witten, ``Nonperturbative Superpotentials in String
Theory,'' hep-th/9604030.}
\lref\kss{S. Kachru, E. Silverstein, and N. Seiberg, ``SUSY Gauge
Dynamics and Singularities of 4d N=1 String Vacua,'' hep-th/9605036.}
\lref\ftheory{C. Vafa, ``Evidence for F-theory,'' hep-th/9602022\semi
D. Morrison and C. Vafa, ``Compactifications of F-theory on Calabi-Yau
Threefolds I, II,'' hep-th/9602114,9603161.}  
\lref\sixmap{M. Bershadsky, K. Intriligator, S. Kachru, D. Morrison,
V. Sadov, and C. Vafa, ``Geometric Singularities and Enhanced Gauge
Symmetry,'' hep-th/9605200.}
\lref\ADS{I. Affleck, M. Dine, and N. Seiberg, ``Dynamical Supersymmetry
Breaking in Supersymmetric QCD",  
{\it Nucl. Phys.} {\bf B}241 (1984) 493.}
\lref\edsmall{E. Witten, ``Small Instantons in String Theory,''
{\it Nucl. Phys.} {\bf B}460 (1996) 541, hep-th/9511030.}
\lref\witbag{E. Witten and J. Bagger, ``Quantization of
Newton's Constant in Certain Supergravity Theories'', {\it Phys. Lett.}
{\bf B}115 (1982) 202.}
\lref\silvwit{E. Silverstein and E. Witten, ``Criteria for Conformal 
Invariance of (0,2) Models'', {\it Nucl. Phys.} {\bf B}444 (1995) 161.}
\lref\ks{S. Kachru and E. Silverstein, ``N=1 Dual String 
Pairs and Gaugino Condensation'', {\it Nucl. Phys.} {\bf B}463 (1996) 369,
hep-th/9511228.}
\lref\dks{J. Distler, S. Kachru, and E. Silverstein, to appear.}
\lref\svw{S. Sethi, C. Vafa, E. Witten, ``Constraints on
Low-Dimensional String Compactification'', hep-th/9606122.}
\lref\schimm{I. Brunner and R. Schimmrigk,
``F Theory on Calabi-Yau Fourfolds'', hep-th/9606148.}
\lref\asp{P. Aspinwall and M. Gross, ``The SO(32) Heterotic
String on a K3 Surface'', hep-th/9605131.}

\Title{RU-96-73}
{\vbox{\centerline{Singularities, Gauge Dynamics,  and}
	\vskip4pt\centerline{Nonperturbative Superpotentials in String Theory}}}
\centerline{Shamit Kachru and Eva Silverstein 
\footnote{$^\dagger$}
{kachru@physics.rutgers.edu, 
evas@physics.rutgers.edu}}
\bigskip\centerline{Department of Physics and Astronomy}
\centerline{Rutgers University}
\centerline{Piscataway, NJ 08855}

\vskip .3in
We describe a class of 4d N=1 compactifications of the $SO(32)$
heterotic/type I string theory which
are destabilized by nonperturbatively generated
superpotentials.  In the type I description, the destabilizing
superpotential is generated by a one instanton effect or 
gaugino condensation in a nonperturbative $SU(2)$ gauge
group.  The dual, heterotic description involves destabilization due to
worldsheet instanton or $\it half$ worldsheet instanton effects
in the two cases.  A genericity argument suggests that
a (global) supersymmetry-breaking model of Intriligator and Thomas
might be typical in 
a class of string theory models.
Our analysis also suggests that the tensionless strings which 
arise in the $E_8 \times E_8$
theory in six dimensions when an 
instanton shrinks to zero size should, in some cases,
have supersymmetry breaking dynamics upon further compactification to
four dimensions.  We provide explicit examples, constructed
using F-theory, of the two cases
of dynamically generated superpotentials.

\Date{08/96} %replace this line by \draft  for preliminary versions
	     %or specify \draftmode at some point
%\draft

\newsec{Introduction}

A long-standing problem in string theory has been the tremendous degeneracy
of vacuum states.  Although in the case of vacua with extended supersymmetry
this problem appears to persist nonperturbatively, one expects that 
many of the classical ground states preserving 4d N=1 
supersymmetry should be lifted by nonperturbative effects.  
The ultimate hope is of course to find models with nonperturbative 
superpotentials which leave minima with broken supersymmetry and
vanishing vacuum energy. 

One potential source of nonperturbative superpotentials in string
theory is strong infrared dynamics.  Supersymmetric QCD with 
$N_{f} \leq N_{c}-1$
\ADS\
is a prototypical example of a theory where nonperturbative dynamics 
generates a superpotential.  In this paper, we argue that the 
nonperturbative superpotentials of SUSY QCD and pure
SUSY Yang-Mills also make an appearance 
in four dimensional $SO(32)$ string compactifications on
$K3$ fibrations.  
The dynamics of the $N_c = N_f = 2$ theory has already
been used in \kss\ to explain poles in the Yukawa couplings
of conventional SO(32) string compactifications on Calabi-Yau
threefolds, while the $N_f = 0$ theory made an appearance
in an N=1/N=0 dual pair proposed in \ks. 
Here, we use similar ideas to describe type I 
string compactifications and F-theory compactifications where
a destabilizing superpotential is generated by

(a) instanton effects in a nonperturbative $SU(2)$ gauge group with
two massless doublets, or

(b) gaugino condensation in a nonperturbative pure $SU(2)$ gauge theory.

\noindent
Both (a) and (b) can occur at small instanton singularities in the
classical moduli space.  The superpotentials generated in both cases
are nonperturbative in the type I string coupling, and map to
worldsheet instanton and $\it half$ worldsheet instanton effects in
the heterotic string.

The destabilizing superpotentials which we find can be traced to the
presence of singularities, which arise at codimension 1 in the moduli
space of vacua in case (a) and which are $\it frozen$ into the
compactification in case (b).  In the type I description the singularities
involve degenerations of the gauge bundle over one point in each
$K3$ fiber.
Such singularities can be interpreted as requiring the presence of
(wrapped) five-branes in the vacuum, as in \kss. 
In a dual F-theory description, the analogous singularities are given 
by elliptic fourfolds with $A_1$ degenerations of the elliptic fiber
over a surface.
While to a naive perturbative string theorist generically
singular vacua which exhibit a superpotential of type (b) would
have seemed problematic as compactification spaces, 
it is clear in light of recent progress
that such vacua are perfectly sensible, and simply involve
suitable wrapped D-branes.
In fact, in some sense such vacua should be
generic, in that we no longer need to impose conditions
that ensure that the conformal field theory alone (without inclusion
of D-branes) is nonsingular.

In many standard compactifications of heterotic/type I string
theory, one has unbroken
nonabelian subgroups of the perturbative gauge group which are
generically present, sometimes without charged matter.  
It is worth emphasizing here that the phenomena we are studying
do $\it not$ depend on any features of an unbroken perturbative
gauge group.  The nonperturbative gauge groups we are studying arise
purely from local degenerations of the string compactification, and
can 
occur in the absence of any unbroken perturbative gauge symmetry.  In fact
in the heterotic string, they lead to effects which occur
at the level of worldsheet instantons and so are $\it stronger$
(at weak string coupling) than any nonperturbative effects associated
with the perturbative gauge groups.

\newsec{Singularities, Gauge Dynamics, and Tensionless Strings} 

In this paper, we study singularities in heterotic/type I compactifications
on threefolds $M$ which are $K3$ fibrations.  In choosing such a 
compactification, one must also specify a stable, holomorphic vector
bundle $V \rightarrow M$ with suitable properties \oldwit. 

As discussed in \kss, degenerations of $V$ 
which occur at a point in the generic (smooth) $K3$ fiber of $M$ (and hence
occur over a rational curve $C$ in $M$) are equivalent to
some configuration of 
``small instantons'' or D 5-branes \edsmall\ wrapped on $C$.\foot{We are
assuming the $K3$ fibration admits 
section(s) and choosing one so we can discuss
the base of the fibration $C$.}   
We focus here on configurations involving a single small
instanton.
By taking a limit where $C$ is large, one can $\it derive$ the
spectrum of the four-dimensional theory from 
the spectrum of the six-dimensional compactification on the
fiber $K3$. The 4d theory is a twisted version of
the six-dimensional theory compactified on $C$, where the
twist preserves 4d N=1 supersymmetry.
The zero modes of the relevant Dirac operators on $C$ 
yield the massless 4d spectrum.  

The spectrum in six dimensions, for a single small instanton,
consists of $SU(2)$ gauge theory with 32 doublets of $SU(2)$
\edsmall.  
A 4d N=1 $SU(2)$ gauge multiplet survives the reduction of
the 6d N=1 gauge multiplet to four dimensions. 
Determining the spectrum  
of charged matter fields which survive involves more
calculation: 
2 $\times$ rank(V) of the doublets couple to the connection on
$V$ and $V^{*}$ while the rest couple to a trivial gauge 
connection.   
Any holomorphic bundle $V$ splits as a direct sum of
line bundles $\sum_i {\cal O}(n_i)$ when restricted to
a rational curve $C$.\foot{${\cal O}(n)$ denotes
the line bundle with $c_1 = n$ over $C$.} 
As discussed in \kss, the
massless spectrum which survives at the singularity
depends crucially on the splitting type of 
$V$ when restricted to $C$ -- the surviving doublets
correspond to elements of $H^{0}({\cal O}(n_i-1))$ and
$H^{0}({\cal O}(-n_{i}-1))$.

In \kss, we studied a bundle which split as 
${\cal O}(2)\oplus{\cal O}(-2)\oplus{\cal O}(0)$.  The
resulting spectrum, combined with a tree-level
superpotential, dynamically generated a pole in
the Yukawa coupling of charged generations. 
In the remainder of this section, we describe
in general terms two classes of singularities of the 4d N=1 vacua
which lead to a destabilizing superpotential.  In the next
section, we discuss in some detail examples (derived from
F-theory) of both types. 

\subsec{Instanton Destabilization}

We expect one type of nonperturbative
superpotential to be generated in the following circumstances.
Suppose there is a 
singular locus at codimension one in the moduli space of
vacua, where 
$V$ is singular over a point on the generic
$K3$ fiber.  Furthermore, suppose that  
$V$ splits 
as 
\eqn\instde{V\vert_{C} ~=~{\cal O}(1) \oplus
{\cal O}(-1) \oplus {\cal O}(0) \oplus \dots}
where $\dots$ denotes more trivial factors.
At such singularities, one obtains an extra nonperturbative 
$SU(2)$ gauge
group with two massless doublets.  As described many years ago 
\ADS, instanton effects destabilize the moduli space of 
classical vacua of this theory.
If we denote by $\Lambda$ the scale of the $SU(2)$,
and by $V_{12}=\epsilon_{\alpha\beta}d_{1}^{\alpha}d_{2}^{\beta}$
the gauge invariant combination of the $SU(2)$ doublets $d_{1,2}$,
then the nonperturbative superpotential is:
\eqn\instsup{W_{dyn} ~=~{\Lambda^{5} \over {V_{12}}}~.}  
Because 
\eqn\gaugemap{\sqrt{\alpha'}\Lambda \sim e^{-R^{2}/{\alpha'}}}
in terms of the radius $R$ of $C$ in the heterotic string
description, we see that \instsup\ is exactly the
sort of superpotential one expects worldsheet
instantons \dsww\ to generate in the heterotic string theory.

A well known theorem of supergravity \witbag\ requires that
a superpotential contribution in an N=1 model $\it must ~have$
a pole at codimension one in the moduli space of vacua if it is not to 
vanish.\foot{This theorem depends on compactness of the moduli
space.  In the cases at hand, one can compactify the moduli
space by adding ``points at infinity'' and one knows that
the relevant poles do not occur at infinity.} 
This theorem has been used in \silvwit\ to
strongly constrain the possible superpotentials in  
N=1 vacua describable as gauged linear sigma models.
In more general models, however, one expects 
to find singular loci where \instde\ occurs: these
are concrete examples of heterotic string vacua 
which are destabilized by worldsheet instantons. 
The required pole in the superpotential is none other than
the pole of \ADS, appearing in \instsup.  This agrees nicely
with the observation in \silvwit\  
that worldsheet instanton superpotentials can be 
generated when 
a singularity of the gauge bundle coincides with a 
rational curve in the manifold -- in this case the
relevant curve is $C$, and the singularity is the small
instanton fibered over it.

\subsec{Gaugino Condensation}

Another interesting situation, 
which maps to a novel effect in the heterotic
string, is the following.  Suppose one has a model in which, at
$\it generic$ points in the moduli space of vacua,
$V$ is singular over a point in each $K3$ fiber.
Let the splitting type of $V$ over $C$ be trivial, i.e. 
\eqn\gensplit{V\vert_{C} ~=~{\cal O}(0) \oplus {\cal O}(0) \oplus \dots}
Indeed, this is the generic splitting type for a holomorphic
vector bundle on ${\bf P}^1$ (with $c_1=0$).  
Then one expects a pure $SU(2)$ N=1 Yang-Mills theory to survive 
as the nonperturbative spectrum at the singularity, 
from the wrapped fivebrane analysis of \kss.  
One would have been uncertain about admitting that such 
generically singular compactifications were physically sensible before
the realization that singular gauge bundles can have a concrete 
physical description as type I fivebranes \edsmall.
It is now
clear that such generically singular models simply correspond to 
N=1 vacua where (wrapped) fivebranes are generically
present.\foot{This is very similar to the realization of
compact D-manifolds \bsv\ provided by F-theory \ftheory.}

The strong infrared gauge dynamics of the $SU(2)$ gauge group
induces gaugino condensation and a superpotential
\eqn\gaugsup{W ~=~\Lambda^{3}}
where $\Lambda$ is the scale of the $SU(2)$ theory. 
Although in global supersymmetry this would be a 
harmless constant contribution to the superpotential, in 
the string theory
$\Lambda$ is dynamical (as in \gaugemap) and
\gaugsup\ destabilizes the vacuum.

It is worth remarking that in the $SU(2)$ Yang-Mills theory, one can think
of the superpotential \gaugsup\ as being generated by a 
$\it half$-instanton effect.  A choice of phase has been made in
\gaugsup, and there are actually $\it two$ vacua with 
$W = \pm \Lambda^{3}$, reflecting an unbroken $Z_2$ discrete
shift symmetry of the $\theta$ angle.     
This unbroken $Z_2$ must map to an unbroken shift symmetry for
the axion partner of the Kahler modulus controlling the size of 
$C$ in the heterotic string.  Therefore, we conclude that in 
the heterotic string the superpotential \gaugsup\ will appear
to be a $\it half$ worldsheet instanton effect!

Of course it is difficult to see how a half-instanton
effect can arise from maps of the worldsheet into
the compactification manifold.\foot{In the case
of gaugino condensation in N=1/0 dual pairs obtained as orbifolds of 
N=2 models \ks, the origin of these effects is clear.
The orbifold group is a $Z_2$ which acts on the base of the fibration
by identifying antipodal points.  Then a worldsheet can consistently
wrap ``half'' the curve.}  However recall that the superpotential
\gaugsup\ arises from the two-doublet case by integrating
out the massive doublets.  In other words, add to \instsup\ 
a coupling to a bundle modulus $m$ which is turned on
at generic points in the moduli space where \gensplit\ applies.  
Then 
integrate $V_{12}$ out from the full superpotential,
\eqn\masssup{W_{Tot}=mV_{12}+{\Lambda_H^5\over{V_{12}}}.}
One finds $V_{12}=\pm\sqrt{{\Lambda_H^5\over m}}$ and
$W=\pm 2\sqrt{m\Lambda_H^5}\sim \Lambda^3$, reproducing
\gaugsup.  Presumably then, the sigma model on the
heterotic side is computing the superpotential
\masssup, which involves only integral instanton
effects plus mass terms.  

\subsec{Singlet Couplings and K3 Fibrations}

As discussed briefly in the previous subsection, in general
we expect that varying some of the vector bundle moduli (while fixing the
moduli that would remove the singularity completely) 
can change the splitting type
of the vector bundle. 
For generic VEVs of bundle moduli,
we expect trivial splitting \gensplit, while at nonzero codimension
one can obtain various nontrivial splittings, with
\eqn\nextmost{V \vert_{C} = {\cal O}(1) \oplus {\cal O}(-1) \oplus
{\cal O}(0) \oplus \dots}
the next most common.  
This gives rise to an $SU(2)$ gauge theory with singlet couplings to 
some number of $SU(2)$ doublets, where the relevant singlets are the
bundle moduli which do not remove the singularity.              
This is precisely the
structure of a model of dynamical (global) supersymmetry breaking
considered by Intriligator and Thomas \IT.

They considered supersymmetric QCD with $N_F=N_C$ 
plus singlets coupling to the flavors in such a way
as to generically give mass to them.  Specializing
to SU(2), we have 4 doublets $d_i,i=1,\dots,4$ which
combine to give gauge invariant coordinates 
$V_{ij}=\epsilon_{\alpha\beta}d_i^\alpha d_j^\beta$.
In addition consider singlets $S^{ij}$ coupling
to the other fields in 
the superpotential 
\eqn\supint{W_{IT}=S^{ij}V_{ij}+\lambda(Pf(V)-\Lambda^4).} 
Integrating out $S$ in global supersymmetry, 
we see that supersymmetry is broken,
since the point $V_{ij}=0$ has been removed from
the moduli space by quantum effects.  In supergravity, 
the condition becomes more complicated; instead
of ${\partial W\over {\partial S}}=0$, it
becomes ${\partial W\over {\partial S}}+K_{,S}W=0$
where $K$ is the Kahler potential.  So unfortunately
supersymmetry breaking is not automatic in supergravity. 
Also, in the string theory setting, $\Lambda$
depends on a dynamical field (the size of the 
${\bf P}^1$ base of the K3 fibration).  
Therefore in order to fix this modulus we would
require for example a product gauge group which
could arise from multiple fivebranes wrapping the base
(which is consistent since there is typically a family
of base ${\bf P}^1$s).  

Note that the linear term in $S$ was crucial above.\foot{We
thank N. Seiberg for emphasizing this point.}
One could also consider the situation where this
is replaced by a nontrivial function $f(S)$.  
This would leave a supersymmetric vacuum
at $f'(S_{min})=0$, and the superpotential
would no longer force the singularity to be frozen in.

%The reader may be puzzled as to why supersymmetry
%breaking through the F term for $S$ is advantageous
%given that we are already discussing gaugino
%condensation, which has been much studied in
%the context of supersymmetry breaking in string
%theory following \drsw.  Subsequent study
%revealed difficulties in establishing
%that supersymmetry is broken in the true vacuum
%of such theories, as this depends on the details
%of the moduli-dependence of the potential.
%(this difficulty is present at weak heterotic coupling, where the problem
%has been traditionally studied, as well as at
%the intriguing limit of
%strong coupling, where the overall phenomenology
%seems more promising \strongphen).  
%In one class of such models where supersymmetry
%is clearly broken--the so-called ``no scale'' models--there 
%is a modulus whose VEV is left unconstrained.
%Therefore
%it is encouraging to find the Intriligator/Thomas
%mechanism for supersymmetry breaking, which does not
%automatically leave a no-scale type modulus, 
%as a likely situation in generic string models.
%!!CHECK NO SCALE--IS THIS TOO STRONG?

\subsec{$E_8 \times E_8$ Tensionless Strings}

One of the most interesting developments of string duality has 
been the realization that tensionless noncritical strings
become the dominant low energy modes in some corners of
moduli space \edcomments.  In particular, this occurs in the
$E_8 \times E_8$ heterotic string when an instanton
shrinks to zero size \ganhan\seiwit. 

Consider
the $E_8 \times E_8$ string on a $K3$ 
fibration $M$ with a vector bundle $V$ 
which develops a small instanton singularity in
the generic $K3$ fiber.  On further compactification on
an $S^1$, this theory becomes equivalent 
to the $SO(32)$ theories we have been studying,
also compactified on $M \times S^{1}$.  This
equivalence (based on T-duality)
depends on the use of Wilson lines--one must check
in a given example that these are compatible with $V$.

In the cases of \S2.1 and \S2.2, the 
fibered small instanton leads to vacuum destabilization
for the $SO(32)$ strings.  By the equivalence mentioned
in the previous paragraph, 
this implies that the three-dimensional $E_8 \times E_8$
vacua on $M \times S^1$ also have no stable supersymmetry
preserving vacuum.  This indicates that the dynamics of 
the tensionless string in $\it four$ dimensional 
N=1 vacua can itself
break supersymmetry.

As a corollary of this discussion, one also predicts
that there should be a family of different $E_8 \times E_8$
tensionless strings in four dimensions.  For each possible 
splitting of $V$ on the base curve (over which it is generically
singular), one should get a different tensionless string theory
in four dimensions.  It is important in the preceeding discussion
that the 6d tensionless strings are effectively being compactified
on $C$ which has $\pi_{1}(C) = 0$.  This means that there are no
winding modes of the tensionless strings (which would have appeared
as particles in four dimensions), so nothing will replace the noncritical
string as the most relevant low energy degree of freedom.

\newsec{F-theory Examples}

The discussion in section 2 was general; it is of obvious
interest to get control over specific examples of the
above phenomena.  
The N=1 heterotic models that have been the focus of 
previous studies have been realized as
gauged linear sigma models on the worldsheet \phases\dk.  
The superpotential vanishes in such models, so the conditions
under which destabilization is possible are very 
limited \silvwit.\foot{Destabilization would require the presence of 
a ``twisted singlet'' (i.e. a singlet whose VEV would describe
a deformation away from the linear
sigma model moduli space) with linear coupling to the other
singlets in a nontrivial superpotential with the correct
singularity structure.}
One way to construct more generic models, which do admit the 
phenomena of \S2, is to use the duality between heterotic models and
F-theory (as was done in \witsup).\foot{For  
a discussion
of an analogous question of superpotentials
on 3-branes in six dimensions,  where it is 
relatively simple to construct examples as gauged linear sigma models
on the  
heterotic side, see \dks.}

One can derive F-theory/heterotic dualities for four-dimensional N=1
theories by starting with the accepted dualities in higher dimensions
\ftheory\ and applying the adiabatic argument.  This leads one to 
believe that F-theory compactified on an elliptic fourfold $X$ which
has as base a rationally ruled threefold $B = {\bf P}^{1}\rightarrow B^{'}$
should be dual to some heterotic string compactification on the
elliptically fibered Calabi-Yau $Y = T^{2} \rightarrow B^{'}$.
Discussions of such dualities have appeared recently in
\witsup\svw\schimm. 

One peculiar feature of these compactifications is that the F-theory
side requires the presence of $\chi(X)/24$ threebranes for consistency.
These must map to fivebranes, wrapped around the $T^{2}$, in the dual
heterotic compactification on $Y$.
In this section, we will be using purely local considerations which
should not be affected by the presence of such branes -- we can
move them away from the relevant singular locus so our arguments
should be valid also in cases when no branes are present.\foot{As
explained in \svw, it is not clear what the precise situation is
with regard to the need for threebranes in cases when $X$ is
generically singular, which are the cases of interest for us.} 

In the following two subsections we will present examples of F-theory
compactifications on elliptic fourfolds $X_1$ and $X_2$ which are dual,
by the adiabatic argument, to type I/heterotic compactifications
on threefolds $Y_{1,2}$ which exhibit the phenomena of \S2.1 and \S2.2

\subsec{An Example of Instanton Destabilization}

The fourfold $X_1$ can be described as follows. 
It is given as a hypersurface in a toric variety which is of the rough
form $C^{9}/(C^{*})^{4}$.
There are homogeneous coordinates $r,s,u,v,p,t,x,y,q$ and the
$C^{*}$ symmetries $(\lambda,\mu,\nu,\rho)$ act on these
coordinates with exponents given in the table below:

\medskip
\halign{\indent\hskip 1.0in #
\qquad\hfil&\hfil#\quad\hfil&\hfil#\quad\hfil&
\hfil#\quad\hfil&\hfil#\quad\hfil&\hfil#\quad\hfil&\hfil#\quad
\hfil&\hfil#\quad\hfil&\hfil#\quad\hfil&\hfil#\quad\hfil\cr
&$r$&$s$&$u$&$v$&$p$&$t$&$x$&$y$&$q$\cr
$\lambda$&$1$&$1$&$6$&$0$&$16$&$0$&$48$&$72$&$0$\cr
$\mu$&$0$&$0$&$1$&$1$&$4$&$0$&$12$&$18$&$0$\cr
$\nu$&$0$&$0$&$0$&$0$&$1$&$1$&$4$&$6$&$0$\cr
$\rho$&$0$&$0$&$0$&$0$&$0$&$0$&$2$&$3$&$1$\cr
}
\noindent
\medskip
This manifold is a fibration of the elliptic fibration over $F_{4}$ over an
additional $\bf{P}^{1}$.  Because the elliptic fibration over $F_{4}$ is dual
in six dimensions to the type I theory \ftheory, this fourfold should
be dual by the adiabatic argument 
to some N=1 compactification of the $SO(32)$ strings.
In fact, since the base of the elliptic fibration 
$B$ is a $\bf{P}^1$ fibration over $F_6$, 
the F-theory compactification should be dual to a heterotic/type I
compactification on the elliptic fibration over 
$F_6$.

$X_{1}$ has the form of a fibration $T^{2} \rightarrow \bf{P}^{1} 
\rightarrow 
\bf{P}^{1} \rightarrow \bf{P}^{1}$, where the coordinates on the three $P^{1}$s
are
\eqn\coords{z_{1} = r/s, ~~z_{2} = v/u, ~~z_{3} = p/t~~.}
One can imagine taking the $\bf{P}^1$ with homogeneous coordinates $r,s$
to be very large.  In this limit, one obtains a six-dimensional
theory dual to the type I compactification on $K3$.  

As we will show below, at complex codimension one in the moduli space
of vacua, one can see that $X_1$ develops an $I_2$ singularity
(in Kodaira's classification) at $v=0$ which looks (in the six-dimensional
limit) like the F-theory description of a single small instanton
\asp\sixmap, and gives rise to a nonperturbative $SU(2)$ gauge group.
Upon further fibration to four dimensions on the last $\bf{P}^{1}$, which
we call $C$, this yields an F-theory model dual to a type I/heterotic
theory which develops a small instanton in the generic $K3$ fiber at
codimension one.  This is equivalent to a type I vacuum with a D 5-brane
wrapped around $C$.

In addition to the nonperturbative $SU(2)$ gauge group, some of the
charged matter associated to the D 5-brane must also survive the
reduction to four dimensions.  Since the singularity is at complex
codimension one, the simplest possibility is that precisely two doublets
$d_{1,2}$ survive the reduction.  We will argue that this is the case
below.  Hence, we will find ourselves with an example of precisely
the situation described in \S2.1.

These results are most evident if one examines the Weierstrass form
for $X_1$. 
$X_1$ is defined by an equation
\eqn\weierI{y^{2} = x^{3} + q^{4} x \sum_{i=0}^{6} p^{i}t^{8-i}
f_{96-16i,24-4i}(z_{1},z_{2}) + q^{6} \sum_{j=0}^{9} p^{j}t^{12-j}
g_{144-16j,36-4j}(z_{1},z_{2})} 
where the subscripts on the $f$s and $g$s indicate the charges 
under the first two $C^{*}$ actions.

Expanding $f$ and $g$ in $v$, we find
\eqn\f{f\sim a_0 p^6t^2q^4+a_2p^5v^2u^2t^3q^4H_4(r,s);}
\eqn\g{g\sim b_0 p^9t^3q^6+b_2p^8v^2u^2t^4q^6\tilde H_4(r,s)}
where $H_4(r,s)$ and $\tilde H_4(r,s)$ are fourth-degree
polynomials in $r$ and $s$, the coordinates on the
base ${\bf P}^1$.
To understand the singularity structure, we
must plug these into the equation 
\eqn\disc{\Delta = 4f^{3} + 27 g^{2}}
for the discriminant.  Then we see that for
\eqn\locus{4a_0^3=-27b_0^2,} 
\eqn\discI{\Delta\sim v^2t^7(H_4(r,s)+\tilde H_4(r,s)).}
Thus at the codimension one locus \locus, there
is a Kodaira type $I_2$ singularity, corresponding to
enhanced $SU(2)$ gauge symmetry.  Since the singularity
is localized in $z_2=v/u$, this gauge symmetry enhancement
is nonperturbative on the heterotic side \asp\sixmap.  

It is important to understand not only the codimension
of the singularity, but the precise matter content
charged under the enhanced $SU(2)$
that survives in four dimensions.  There is no
direct correlation between the codimension and
the matter content in $4d$ $N=1$ supersymmetry
because of the possibility of a tree-level
superpotential.  

Let us pause to explain this
in more detail.  In six dimensions, each small
instanton is equipped with matter in the ${\bf (32,2)}$ of
$SO(32)\times SU(2)$.  Upon reduction on the
base ${\bf P}^1$, the $4d$ spectrum
will contain some number $n_d$ of doublets; others
will develop Kaluza-Klein masses.  
The surviving $n_d$ doublets could couple in a nontrivial
tree level superpotential.  Generically this would
start at fourth order in the doublets, or at
{\it quadratic} order in the gauge invariant coordinates
$V_{ij}=d_i^\alpha d_j^\beta \epsilon_{\alpha\beta}$;
away from the singularity, these contributions
would look like {\it mass} terms for singlets.
Therefore codimension alone is not a reliable
guide to the physics.  In order to determine the
presence of a dynamically generated superpotential,
one needs at least two of three classical inputs:
charged matter spectrum, codimension of the singularity,
and tree-level superpotential.  In the previous study
\kss, we computed the first two, which determined enough
of the third to proceed to an understanding of $W_{dyn}$.
In the present case, we have a handle only on the codimension
of the singularity.

In this case we can, however, determine the presence of
the expected dynamical superpotential by invoking the
analysis of \witsup.  This will also provide a connection
between the gauge theory cases we are studying here and the 
discussion provided there. 
In \witsup, it was shown that contributions to the superpotential
in F-theory compactification on $X_1$ correspond to divisors
$D$ in the fourfold with 
$\chi(D,{\cal O}_D) =1$, where
\eqn\arith{\chi(D,{\cal O}_D)=h^{0,0}(D)-h^{1,0}(D)+h^{2,0}(D)-h^{3,0}(D)}
is the arithmetic genus of $D$.  Furthermore, the superpotential
will correspond to a worldsheet instanton effect on the
heterotic side if the divisor $D$ involves only a curve
in the last two ${\bf P}^1$s (the part of the F-theory
manifold that is visible on the heterotic side).\foot{The results 
of \witsup\ were derived and applied in the context of models without
generic singularities.  The model under present discussion has a generic
singularity (related to a $\it perturbative$ heterotic gauge group) at
$z_{3} = \infty$, which is irrelevant to the physics we are discussing.
We are assuming that this extra singularity does not invalidate our
use of the results of \witsup.} 

In our example, there is an obvious such divisor $D$:
\eqn\div{v=0.} 
That this divisor has $\chi(D,{\cal O}_{D})=1$ can be seen as follows
(following \S2.1\ of \witsup).  First note that since
$v$ is the only polynomial in the projective coordinates
with charges (0,1,0,0), there are no deformations of this
divisor.  Hence, $h^{3,0}(D)=0$.
One can then use the Lefschetz hyperplane theorem to fix
$h^{1,0}(D)$ and $h^{2,0}(D)$.  This theorem states that
for $D$ a divisor in $M$ determined by the vanishing
of a section of a positive line bundle on $M$,
there is an isomorphism between the integral cohomology
of $M$ and that of $D$, up to (but not including) the
middle dimension.  In our case, since it has positive
charge, $v$ is a section of a positive line bundle,
so the theorem applies to the divisor \div.
Similarly, the theorem applies to the Calabi-Yau
fourfold $X_1$.  Since $h^{1,0}(X_1)=h^{2,0}(X_1)=0$
follows from the fact that $X_1$ is simply
connected and has $SU(4)$ holonomy
instead of a proper subgroup, 
we learn
that our divisor \div\ also has $h^{1,0}(D)=h^{2,0}(D)=0$.
Putting it together, the only contribution to $\chi(D,{\cal O}_D)$
is from $h^{0,0}(D)$, yielding $\chi(D,{\cal O}_D) = 1$.

Therefore the locus \div\ determines a contribution
to the superpotential, which moreover corresponds
to a worldsheet instanton on the heterotic side wrapping
the base of the $K3$ fibration $Y_1 = T^{2} \rightarrow
F_6$.  Let us denote the
Kahler modulus of this ${\bf P}^1$ as $T$.  The superpotential will
contain, in addition to the classical piece
$e^{-T}$, holomorphic dependence on the other moduli.
As discussed in \witbag, the
superpotential must have a pole at codimension one on the
space of chiral superfields, assuming that space is compact.
Therefore, there must be codimension one singularities
in F-theory at which the superpotential can diverge.
In our problem, we have just this structure.  As
explained above, at codimension one in the moduli
space of vacua, the elliptic
fiber has an $A_1$ degeneration on the divisor
\div.  So we have
a superpotential at the one worldsheet instanton level
which blows up at the locus with enhanced non-perturbative
$SU(2)$.  This is just the situation discussed in \S2.1\ for
$SU(2)$ with two doublets.  

\subsec{An Example of Gaugino Condensation}

We can similarly present an F-theory compactification on a fourfold
$X_2$ which is dual to the situation envisioned in \S2.2,  
though with somewhat less control given the
generic singularity and a strange feature which will
soon become apparent.
The toric data for  
$X_2$ is given in the table below:

\medskip
\halign{\indent\hskip 1.0in #
\qquad\hfil&\hfil#\quad\hfil&\hfil#\quad\hfil&
\hfil#\quad\hfil&\hfil#\quad\hfil&\hfil#\quad\hfil&\hfil#\quad
\hfil&\hfil#\quad\hfil&\hfil#\quad\hfil&\hfil#\quad\hfil\cr
&$r$&$s$&$u$&$v$&$p$&$t$&$x$&$y$&$q$\cr
$\lambda$&$1$&$1$&$3$&$0$&$11$&$0$&$32$&$48$&$0$\cr
$\mu$&$0$&$0$&$1$&$1$&$4$&$0$&$12$&$18$&$0$\cr
$\nu$&$0$&$0$&$0$&$0$&$1$&$1$&$4$&$6$&$0$\cr
$\rho$&$0$&$0$&$0$&$0$&$0$&$0$&$2$&$3$&$1$\cr
}
\noindent
\medskip
$X_2$ is a hypersurface in this toric variety with Weierstrass form
\eqn\weier{y^{2} = x^{3} + q^{4} x \sum_{i=0}^{5} p^{i}t^{8-i}
f_{64-11i,24-4i}(z_{1},z_{2}) + q^{6} \sum_{j=0}^{8} p^{j}t^{12-j}
g_{96-11j,36-4j}(z_{1},z_{2})} 
with the same notation as in \weierI.
By the adiabatic argument, it should be dual to a type I/heterotic
compactification on $Y_2 = T^2 \rightarrow F_3$.

It follows from the data defining $X_2$ that none of the $f$s
can contain a term independent of $z_{2}$ (i.e. proportional to $v^{0}$)
and only one, $f_{9,4}$, can contain a linear term in $z_{2}$ (which
is then independent of the other coordinates).  Similarly, none of the
$g$s can contain constants or terms linear in $z_{2}$, but quadratic 
terms in $z_{2}$ are allowed.  This means that the discriminant
is of the form 
\eqn\disctwo{\Delta = z_{2}^{3}t^9({\rm const} + O(z_{2}))}
expanded around $z_{2}=0$.  From \ftheory\sixmap\ one sees that this
is a type III singularity in Kodaira's classification, 
which yields an $SU(2)$ gauge
group.
This means that there is a small instanton in the generic $K3$ fiber
(fibered over the ${\bf P}^{1}$ with coordinate $z_1$) in the
dual heterotic string description.
\foot{It is actually an $I_2$ singularity at $z_2=0$ 
which was found to be
dual to generic small instantons in \asp\sixmap.  The type III
singularity is to be physically thought of as a special (more
degenerate) case of $I_2$ which still yields the same gauge group. 
Imposing type III instead of $I_2$ gives a proper submoduli space
of the small instanton moduli space .}

Similarly to the case of $X_1$, here we have a divisor
$v=0$ with no deformations.  Again the Lefschetz theorem
applies, and using the isomorphism of the cohomology to
that of the fourfold, $\chi(D,{\cal O}_D)=1$ 
and there will be a superpotential.
Here the locus $v=0$ is a frozen in singularity, so
we expect the superpotential to diverge.  This is
consistent with the following superpotential:
\eqn\frozsup{W=AV_{12}^k+{\Lambda^5\over{V_{12}}}}
for some $k$ (the presence of the first term is necessitated by
the fact that the singularity is frozen in).  
If $k=1$ we recover the gaugino condensation
superpotential of \S2.2.  Unfortunately, this
is difficult to determine here.  In any case,
integrating out $V_{12}$ gives a superpotential which
has fractional instanton number.

There are some strange features of this model which make its
dual interpretation as a type I vacuum somewhat hazy.
In particular, in addition to the small instanton singularity,
the gauge symmetry at $z_{3} = \infty$ has been enhanced to
$F_{4}$.  While this $F_4$ can be described as a perturbative
gauge symmetry in the $E_8 \times E_8$ heterotic string dual 
\ftheory\asp\sixmap\ (which one would obtain by fibering the
$E_8\times E_8$ string on $K3$ with (16,8) instantons in the
two $E_8$s over the last $\bf{P}^{1}$), it is not a symmetry
which one expects to arise perturbatively in the type I or
$SO(32)$ heterotic string description. 

If this subtlety with the gauge group at $z_{3} = \infty$ does
not affect the interpretation of the $A_1$ singularity, then this
F-theory model is (dual to) an example of the sort described in
\S2.2: A small instanton in the generic $K3$ fiber of the
type I / heterotic compactification has yielded a pure $SU(2)$
gauge group in four dimensions.  Hence, the vacuum is destabilized
by gaugino condensation / half worldsheet instantons.

\newsec{Conclusions}

Many supersymmetric $SO(32)$ string compactifications, including those with no
interesting dynamics originating from perturbative gauge groups,
should be lifted by nonperturbative effects originating in nonperturbative
gauge groups.  While the examples of \S3 were given in the framework of
F-theory, it is clear that the phenomena of \S2 are far more general and
will occur in heterotic models without any (at least presently known)
candidate F-theory dual.

It would be interesting to explore heterotic/type I models with 
such nonperturbative superpotentials more directly, perhaps using
techniques appearing in \dks.

A complementary approach to the study of N=1 models
with superpotentials will appear in \berkooz, where a particular
type I orbifold is studied in great detail.

\centerline{\bf{Acknowledgements}}

We would like to thank T. Banks, J. Distler, D. Morrison, C. Vafa, and
especially N. Seiberg for
helpful discussions.
This work was completed while the authors were 
enjoying the hospitality of the Aspen Center for
Physics.

\listrefs
\end